\documentclass[epj]{svjour}
\usepackage{psfig,cite,here}
\begin{document}
\title{Competition, efficiency and collective behavior in the
``El Farol'' bar model}
% Complexity of complicity in complicated systems.
\author{{M. A. R. de Cara}\and{O. Pla}\and{F. Guinea} \\
\email{angeles@quijote.icmm.csic.es}
}
\institute{Instituto de Ciencia de Materiales\\
Consejo Superior de Investigaciones Cient\'\i ficas \\ 28049 Madrid SPAIN}
\abstract{The {\bf El Farol} bar model, proposed to study the dynamics of
competition of agents in a variety of contexts (W. B. Arthur, Amer. Econ.
Assoc. Pap. and Proc. {\bf 84}, 406 (1994)) is studied. We characterize in
detail the three regions of the phase diagram (efficient, better than random and
inefficient) of the simplest version of the model (D. Challet and
Y.-C. Zhang, Physica A, {\bf 246}, 407 (1997)). The efficient region is shown to
have a rich structure, which is investigated in some detail. Changes in the
payoff function enhance further the tendency of the model towards a wasteful
distribution of resources.}
\date{\today}
\maketitle

\section{Introduction}

In recent years there has been a growing interest in understanding the
dynamics of systems of interacting individuals with
competing goals (frustration).
Simple rules for the behavior of the individuals may lead to
unexpected properties in the behavior of the collectivity. These rather
general premises can apply to problems in different fields,
like economy~\cite{SantaFe}, ecology~\cite{May} or physics~\cite{algo}.

To illustrate these facts Brian Arthur introduced what he called
\textsl{``El Farol''} bar problem (EFBP)~\cite{arthur}. $N$ individuals 
decide, at each time
step, to go to a bar or to stay at home. The bar is enjoyable only if 
the attendance does not surpass some critical number, that can be thought of
as some kind of \textsl{comfort capacity}. But each individual does not
know beforehand what is going to happen. To be able to make the decision for
the next time step the individuals (which we will call \textsl{agents} in the
following, as in previous literature of this model) are provided each one with
a set of strategies. Using these strategies,
and the knowledge of what has happened in the
portion of the history that they can recall, the agents take decisions.

D. Challet and Y.-C. Zhang~\cite{CZ97} have given a precise set of rules
which determine the model. The two possible choices, going to the bar or 
staying at home, are represented by 0 and 1. A choice is successful if the
agents which make it are in the minority (comfort capacity = 50\%). The
outcome of a given simulation
is represented by a series of 0's and 1's which characterize the 
successful choices at each time step. Each agent uses a fixed set of $s$
strategies, taken at random from the pool of all possible strategies.
Strategies use the full information of the $m$ previous outcomes to
decide the next move. As there are $2^m$ possible combinations
of past events, the
number of strategies is $2^{2^m}$. After each event, the agents
update the score of their set of strategies. The gain made by
the successful strategies can either be a fixed constant, or depend
on the size of the group formed at that time step. In the simplest version
of the model, one point is assigned to each successful strategy. 
When an agent has two or more strategies with the same score, 
one of them is picked at random. This choice of payoff is
the one discussed
in detail below. The model is defined by the
three parameters: $N$, the number of agents, $m$, the number of time 
steps used by each strategy in determining the next best move, and $s$,
the number of strategies available to each agent.
Extensions to other payoff schemes, similar to
those used in~\cite{CZ97,CZ98,Z98} are also mentioned. Note that the original 
work~\cite{arthur} used a much less constrained set of strategies and a
different comfort capacity (60\%).

The model, with the set of rules described above, was investigated 
in~\cite{savit,CZ98}. The authors analyze the mean size 
and the fluctuations of the groups taking
each of the two choices available. It is argued that the
model can be characterized in terms only of the 
combination $\rho = 2^m/N$. The average group size is $\frac{N}{2}$.
The distribution of sizes is symmetrical around this value.
The mean quadratic deviation from the average, $\sigma$, is a measure of
the number of points accumulated by all the agents. This number is 
maximum when the two groups are almost equal, in which case 
$\sigma \sim O ( 1 )$. As function of $\sigma^2/N$ and
$\rho$ three regimes can be distinguished,
as function of the total number of strategies at play~\cite{savit}: 
i) When $\rho \gg 1$, the number of strategies available to the agents
is small, and the value of $\sigma$ approaches the limit expected
when the agents take random decisions,  $\sigma^2/N = 1/4$.   
ii) If $\rho \ll 1$, almost all possible
strategies are in possession of the agents, and their performance is
worst than random, as $\frac{\sigma^2}{N} > \frac{1}{4}$. 
iii) Finally, for $\rho \sim 1$ the agents perform statistically better
than random. The curve of $\frac{\sigma^2}{N}$ versus $\rho$ shows a
minimum. The authors define regime i) as inefficient, as the agents
have little information, and regime ii) as efficient, as agents have
all available information at their disposal.

In section~\ref{s:minor}, we analyze the model defined above, with emphasis
on the structure shown in the efficient region.
Section~\ref{s:interp} presents an interpretation of the results.
Then, section~\ref{s:pay}, we discuss results
obtained by varying the payoff function which determines the choice
of strategies. Section~\ref{s:major} analyzes a seemingly trivial variation of
the model: the majority game, when it becomes preferable to be in the
majority. The final section presents the conclusions.

\section{Minority game.\label{s:minor}}

\begin{figure}[h]
\begin{center}
\mbox{\psfig{file=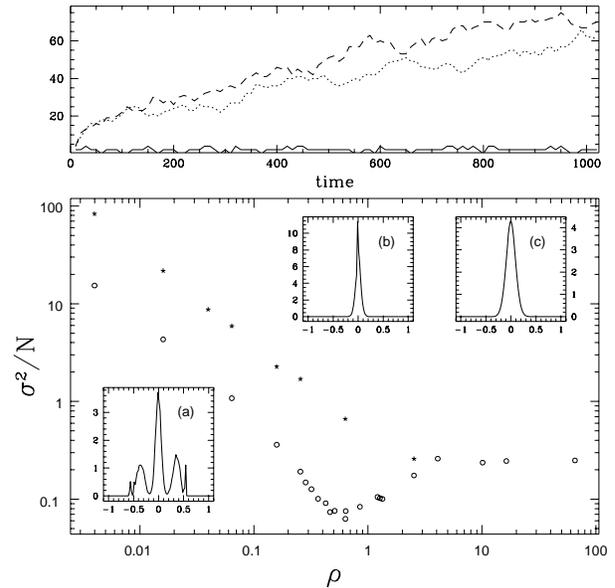,width=8cm}}
\end{center}
\caption[Short]{Different phases found in the EFBP. The lower part shows the 
evolution of $\sigma^2/N$ as function of $\rho$,
circles are for $s$=2 and stars are for $s$=6. The insets show 
histograms of the attendance number in the different
phases, with $N=101$: a) efficient, $m$=2, $s$=2; b) better than random $m$=6,
$s$=2; and c) inefficient $m$=10, $s$=2. 
The top figure shows the difference in punctuation between the maximum scored
and the minimum scored strategies in these three cases: dotted line for (a),
dashed line for (b), and continuous line for case (c).}
\label{fig:trans}
\end{figure}

The transition discussed in~\cite{savit} is displayed in 
fig.~\ref{fig:trans} for $s=2$ and $s=6$. The difference between the
efficient and inefficient regimes is sharper for small 
values of $s$. Each simulation of the model starts from a
history of length $m + 3$ to initialize the scores of the strategies. 
The results shown in the paper are averages over the $2^{m+3}$ 
possible initial conditions defined in this way. In almost all
cases, the system evolves towards a steady state which is independent of
the initial conditions.
\begin{figure}[h]
\centering{\mbox{\psfig{file=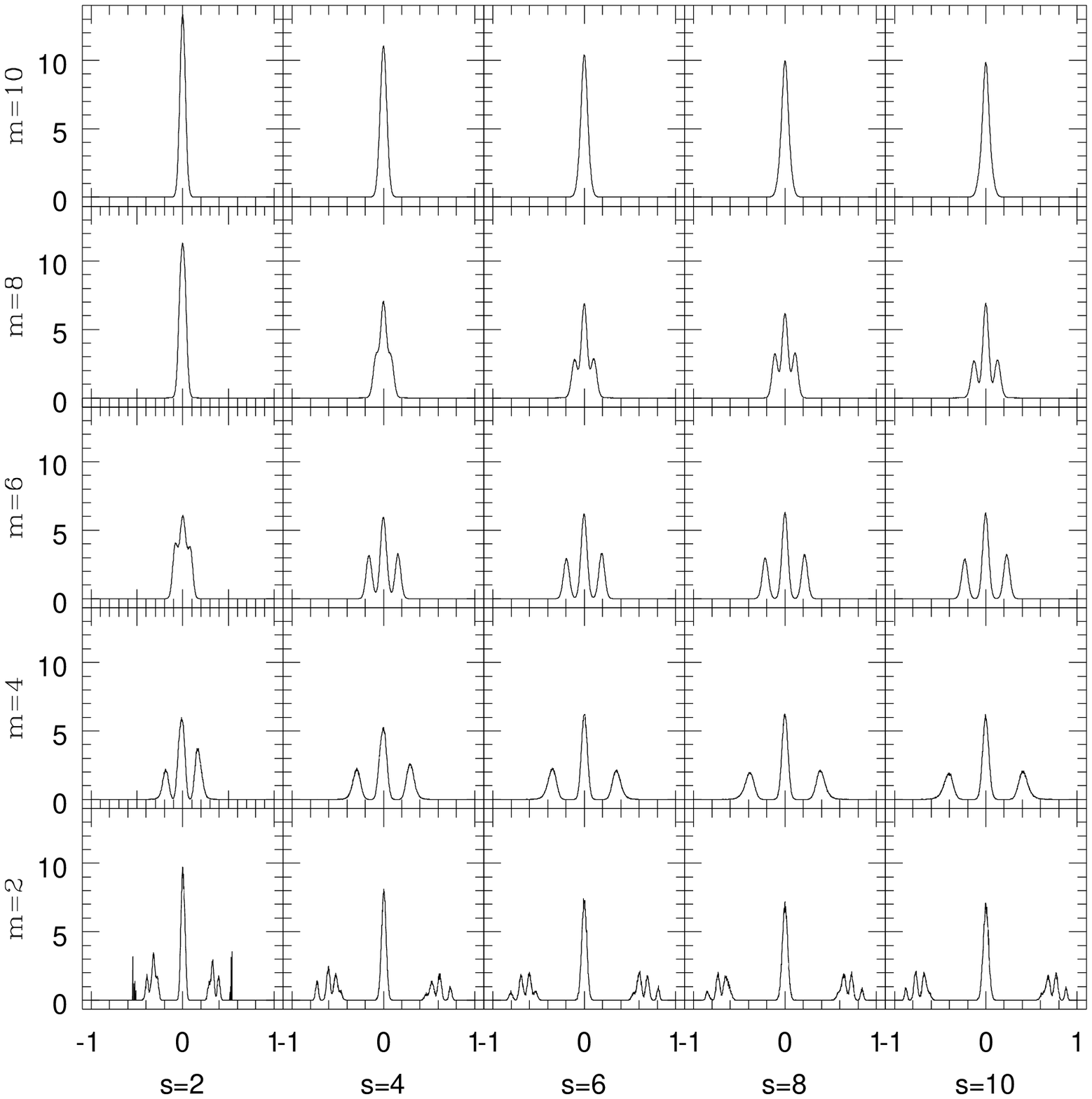,width=8cm}}}
\caption{Attendance numbers distributions for $N=1001$}
\label{peaks}
\end{figure}

\begin{figure}[h]
\begin{center}
\mbox{\psfig{file=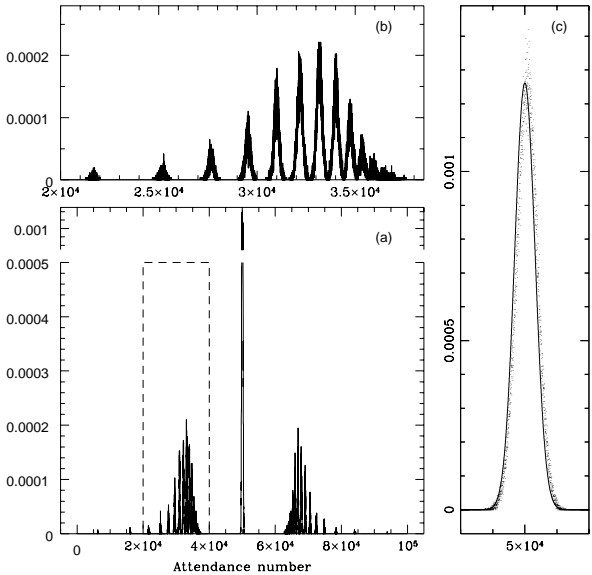,width=8cm}}
\caption{Attendance numbers distribution for $N$=100001, $s$=4,
and $m$=4, normalized in the interval $[0,100001]$.
(a) Full distribution where the y-axis has been truncated in order to
appreciate the spreading of the lateral peaks. (b) Magnification of the
region marked in (a) with dashed lines. (c) Points in the central peak. The
continuous line is a Gaussian, centered at $N/2$, with weight half of the
total distribution and deviation $\sqrt{N}/2$.}
\label{fig:cmil}
\end{center}
\end{figure}

The peaks in the size distributions are always well approximated by
Gaussian functions. The large value of $\sigma$ in the efficient
region is due to the formation of new peaks away from $N/2$.
A pictorial view of this effect is shown in fig.~\ref{peaks}, where
the different regimes are studied by varying $m$ and $s$. The attendances
have been normalized to one in the interval $[ -1 , 1 ]$.
In the range of values of $\rho$ where three peaks can be clearly
resolved, the weight of the central peak is one half of the total,
and the other two peaks include one fourth of the recorded
attendances. The central peak is always well approximated by a
Gaussian of width $\sqrt{N}/2$ (see also fig.~\ref{fig:cmil}), which
corresponds to random choices by the agents.

As one leaves the efficient region, the peaks merge with the 
central one, whose width decreases first and then increases, to reach
the random value for large values of $\rho$. For small values of
$\rho$, the peak structure is very rich, and seems self
similar, as shown in fig.~\ref{fig:cmil}. 

As pointed out in~\cite{savit}, it is somewhat unexpected the poor performance
of the agents when a large amount of information is available. 
It is even more remarkable the rich structure shown in fig.~\ref{fig:cmil},
which shows that the evolution is far from random. This behavior is
also consistent with the existence of non trivial patterns in the time
series, beyond the reach of the agents~\cite{savit}.

\begin{figure}[h]
\centering{\mbox{\psfig{file=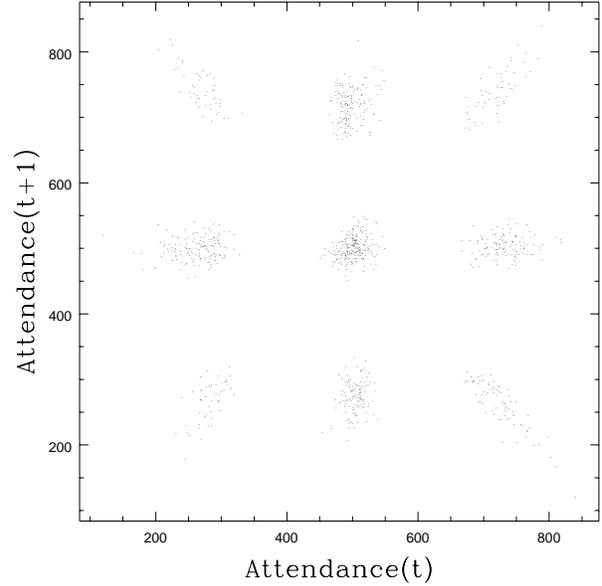,width=8cm}}}
\caption{Attendance in a given group at two successive intervals.
The parameters used are $s=2, m=2$ and $N=1001$.}
\label{fig:despl}
\end{figure}

A plot of the attendances at successive times is shown in fig.~\ref{fig:despl}.
We have chosen the parameters in such a way that the distribution of
attendances shows three separated peaks. 

\begin{figure}[h]
\begin{center}
\mbox{\psfig{file=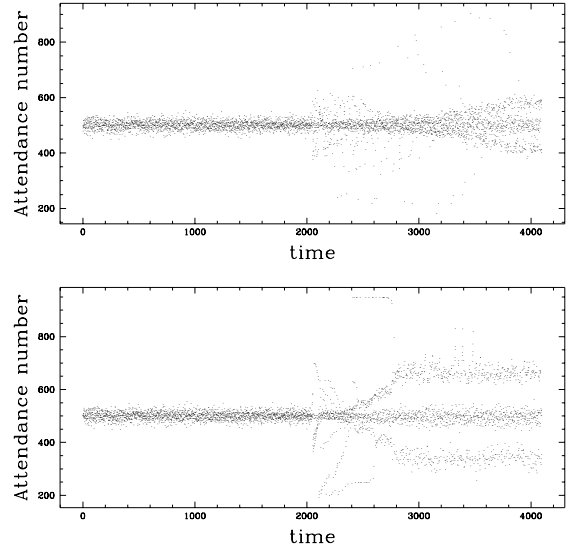,width=8cm}}
\caption{Attendance number versus time for the game in which a transition is
forced from a random game to a minority game (see text). The parameters of
the minority game are: $N$=1001, $s$=4, and $m$=4 (6) for the bottom (top)
graphs.}
\label{fig:azdet}
\end{center}
\end{figure}

We have completed the study the evolution of the different peaks by
analyzing their evolution after an initial series of random choices.
In the time series shown in fig.~\ref{fig:azdet}, the agents make
choices randomly, although their strategies keep updating the scores.
At a given time step (2048), the agents start to use the strategies at
their disposal.  

The peak structure is robust, and develops immediately. As shown in
fig.~\ref{fig:azdet}, the peaks split from the central peak and move to their
positions in the steady state discussed earlier. 

\section{Interpretation.\label{s:interp}}

The results presented in the previous section allow us to gain some
understanding of the complex dynamics of the efficient regime.
In this region, no strategy can stay with the highest score for long.
The repeated use of a given strategy by a significant number of agents
leads to the raise of other strategies, preferably those
more anticorrelated with the one at play. As a result of this, the most
punctuated strategy (the best considered by the agents) has many chances of
making its users to loose. And, eventually, the agents segregate
into anticorrelated groups when some degree of evolution is
incorporated~\cite{johnson}.

For simplicity, we now assume that there are two anticorrelated
strategies, $x$ and $\bar{x}$ which have the highest scores most
of the time. Let us denote $n_x$ and $n_{\bar{x}}$ the number of
agents which have strategy $x$ and $\bar{x}$. We can take
$n_x \approx n_{\bar{x}} = n_{correl}$. We now denote as
$n_{random}$
the number of agents which have neither $x$ nor $\bar{x}$. The choices of
these $n_{random}$ agents can be taken to be at random,
as they are unable to recognize the series which give rise to the
high scores of $x$ and $\bar{x}$.

When strategy $x$ has the highest score, the two groups will have
sizes close to $n_{random}/2 + n_{correl}$ and
$n_{random}/2 - n_{correl}$, respectively. This outcome 
will give no points to $x$, while strategy $\bar{x}$, which would
have lead to the most favorable choice, gains one point.
If the score of $\bar{x}$ remains below that of $x$, the 
process repeats itself. A steady state is reached when the scores
of $x$ and $\bar{x}$ differ by, at most, one point. 
Then, an outcome with two unequal groups of
sizes $n_{random}/2 + n_{correl}$ and
$n_{random}/2 - n_{correl}$ is followed by 
the formation of two groups of similar size, $\approx N/2$.
The fact that there are $n_{random}$ agents acting at random 
implies that these values are the average of Gaussian peaks 
of similar width.

We can estimate the value of $n_{correl}$ from the analysis 
in~\cite{Z98}. We classify the $2^{2^m}$ strategies into
$2^m$ mutually uncorrelated, maximally correlated or anticorrelated
classes. Then, $n_x \approx N/2^m = 1/\rho$.

The previous analysis gives a plausible explanation of the three 
peaks observed throughout most of the efficient region of
parameter space. It can be extended, in a straightforward way,
to the case when the dominant strategies are more than two.
The main new ingredient is that  there are situations in which
two, or more, dominant strategies can have
the same score. Let us imagine that the strategies with the
highest scores are  $x_1, x_2, \bar{x}_1$
and $\bar{x}_2$. Then, at a given instant, the strategy with the
highest score can be $x_1$, $x_2$ ..., but also $x_1$ and $x_2$ 
(or similar combinations) simultaneously. If, in addition,
$x_1$ and $x_2$ lead to the same outcome, the majority
group will be of size $n_{random}/2+ n_{x_1} + n_{x_2}$.
This combination will be, probably, less likely, leading to
lower peaks further away from the average, in agreement with
the findings reported here.

We have checked that there is a trivial case where this analysis
reproduces the observed evolution: $m=2$ and $s=16$, where all 
agents have all strategies. The attendance histograms show 
two sharp peaks at 1 and $N$, and a Gaussian peak with half the weight of the
total distribution at $N/2$, and deviation $\sqrt{N}/2$.

\section{Varying the rewards for the winners.\label{s:pay}}

We now look to the effect of changing the way in which the
different strategies are updated after each outcome. The simplest
modification is to relate the change in the score to the size of
the minority group~\cite{CZ97,CZ98}. In the following, we
assume that the payoff, $\Delta p$, depends linearly with the size, $a$.
If the score is incremented by $a$, strategies which lead
to groups with attendances close to $N/2$ are favored.
If, on the other hand, the score is incremented by $N/2 - a$,
the tendency is the opposite, and strategies which lead to very
small groups are favored.

\begin{figure}[h]
\begin{center}
\mbox{\psfig{file=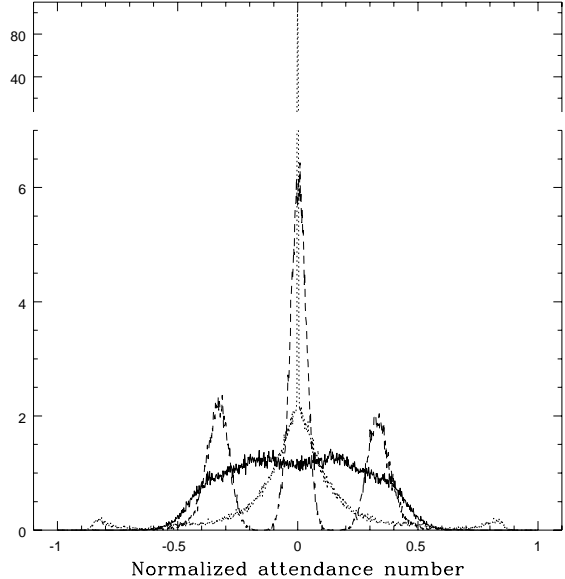,width=8cm}}
\caption{Attendance distributions for $N$=1001, $m$=4, and $s$=4. Dashed
line is for the step payoff, continuous line for $\Delta p = a$, and dotted
line for $\Delta p = N/2 - a$.}
\end{center}
\label{fig:payoff}
\end{figure}

The distributions generated by these two payoff choices
are plotted in figure~\ref{fig:payoff}. The distribution obtained
by the step payoff discussed in the previous section is also
plotted, for comparison.

Contrary to intuition, the two distributions seem to go in the opposite
direction to what the choice of payoff leads to think. It must be noted
that, when the second choice of payoff function is shifted by a
constant, $\Delta p = N/2- a+k$, 
the central peak tends to disappear,
and it is replaced by two peaks at the sides. This result is similar to
other findings with a payoff which also favors small groups~\cite{CZ98}.

We interpret the broad structure for the payoff function $a$ as 
due to the swift shuffle of the highest ranking strategies. Outcomes
with nearly equal groups give rise to large changes in the scores
of the strategies. Thus, long living cycles, of the type described
in the previous section, cannot form. The highest ranking strategy
changes rapidly. As all strategies are in play, groups of many
sizes are generated, despite the fact that the payoff favors 
sizes close to $N/2$.

In the opposite case, with payoff function equal to $N/2 - a$,
we ascribe the large peak at $N/2$ to frequent situations
when many strategies have the same score. This situation is self sustaining,
as, when the two groups are of sizes $N/2$ and $N/2+1$,
there is no change in the scores of the strategies. This is what happens in
half of the possible $2^{m+3}$ initial conditions, and corresponds to the 
delta peak in fig.~\ref{fig:payoff}. The rest of the distribution is a good
average of what happens in the other half of the initial conditions. The shift
of the payoff by a constant described earlier reduces the probability of 
tie-ups, and leads to a double peaked distribution. These peaks displaced
from the center seem, in this case, related to the two peaks
in the step payoff case. It is likely that the evolution of the
model is governed by cycles with a few dominating strategies.

\section{Majority game.\label{s:major}}

We have also studied the majority game, in which the agents prefer to be 
in a overcrowded bar or leave the bar empty. 
The methodology is the same as in the minority game, in which  
the different initial conditions tend to give similar results. Here,
initial conditions may make big changes in the attendance distributions.

Results are trivial (the full majority is attained at all time steps) only when
all agents have all strategies ($s=2^{2^m}$). Even in this case, and
depending on the initial conditions, the group (0 or 1) which obtains the
majority may oscillate in time.

\begin{figure}[h]
\begin{center}
\mbox{\psfig{file=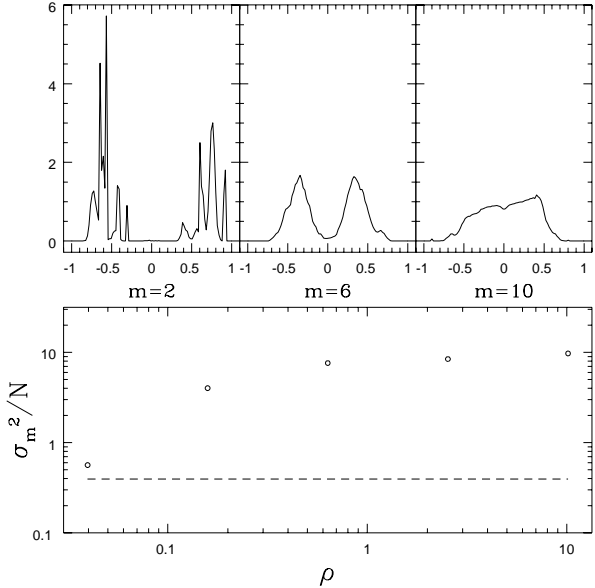,width=8cm}}
\caption{The analogous diagram of figure \protect\ref{fig:trans} for the
majority rule. Here $N$=101 and $s$=4. The dashed line is for 
$\sigma_m = N/2^s$}
\end{center}
\label{fig:mayor}
\end{figure}

The obtained distributions for different values of $m$ and, consequently,
$\rho$, are plotted in fig.~\ref{fig:mayor}.

The particular placement of the fixed points makes that a more convenient
measure of the efficiency should be used. We will use the mean deviation,
$\sigma_m$, calculated around the value $N$ for
the attendance, and shifting the attendances $a$ below $N/2$
to $a + N$. Thus, $\sigma_m$ also gives a measure of the overall
gain made by the agents. In the three plots of attendances, where the 
atendance axis is not folded, the two large peaks near the limits
are not shown. These peaks correspond to limit cycles where
the attendances do not fluctuate.

The relative weight of this peak, for $s=4$, at sufficiently large times,
is 0.56 for $m=2$, 0.078 for $m=6$
and 0.031 for $m=10$. 
The number of agents which are able to coordinate among themselves
and take part in this cycle is, on the average, $N - N/2^s$,
if $s < 2^{2^m}$. Then, the lower limit for $\sigma_m^2/N$ is 
$N/2^{2s}$. This value is also plotted in
fig.~\ref{fig:mayor}.

The relative weight of this peak, which represents the
average number of coordinated agents, converges at sufficiently large times
to 0.56 for $m=2$, 0.078 for $m=6$
and 0.031 for $m=10$. 
It is remarkable that the agents are not too effective in acting
in a coordinated manner. Most initial conditions lead to histories
where the majority group is well below the intuitive natural limit.
This result is consistent with the spin glass features reported in~\cite{savit}.

\section{Conclusions.\label{s:conc}}

As we have seen, the \textsl{El Farol} bar problem has a rich structure.
We have focused mostly on the behavior in the efficient
regime, where most of the strategies are at the disposal of the
agents. As already remarked in\cite{savit}, the model has many features
in common with frustrated systems in statistical mechanics. In particular,
most initial conditions lead to a poor performance of the system as a whole.
The model seems unable to select a pool of strategies such that the
global gain by the agents is maximized. In particular, those agents
which have access to the strategies with the highest scores at a given
moment perform worse than those which do not. The latter play basically
at random, and profit from the unproductive coordination of the players
using the nominally best strategies.

This effect seems to remain when the payoff to the different strategies
is varied. It is also remarkable that the intrinsic frustration of the model
shows up when the agents try to be in the majority. Most initial conditions
lead to evolutions where the agents fail to coordinate among themselves.

Financial support from DGCYT through project no. PB96/0875 
and the European Union through project \hfil \\ ERB4061PL970910 is gratefully
acknowledged.

\begin{acknowledgement}
We would like to thank the helpful comments of
N. Garc\'{\i}a, Enrique Louis, Pedro Tarazona and Yi-Cheng Zhang.
\end{acknowledgement}

\end{document}